# I wanna draw like you: Inter- and intra-individual differences in orang-utan drawings

Marie Pelé[1], Gwendoline Thomas[2], Alaïs Liénard[2], Nagi Eguchi[3], Masaki Shimada[3], Cédric Sueur[4,5]

1 : Anthropo-Lab, ETHICS EA7446, Lille Catholic University, Lille, France

2 : Université Sorbonne Paris Nord - UFR LLSHS, Paris, France

3 : Department of Animal Sciences, Teikyo, University of Science, Uenohara, Yamanashi, Japan

4 : Université de Strasbourg, CNRS, IPHC UMR 7178, Strasbourg, France

4 : Institut Universitaire de France, Paris, France

Corresponding author: Marie Pelé, Lille Catholic University, marie.pele@univ-catholille.fr

**Abstract:** Recently discovered, the oldest human abstract drawing is around 73,000 years. Although the origins of drawing behaviour have remained an enigma to this day, light may be shone on the subject through its study among our closest neighbours, the great apes. This study analyses 749 drawings of five female Bornean orang-utans (*Pongo pygmaeus*) at Tama Zoological Park in Japan. We searched for differences between individuals but also tried to identify possible temporal changes among the drawings of one individual, Molly, who drew almost 1,300 drawings from 2006 to 2016. A classical analysis of the drawings was carried out after collecting quantitative and qualitative variables. Our findings reveal evidence of differences in the drawing style of the five individuals as well as creative changes in Molly's drawing style throughout her lifetime. Individuals differed in terms of the colours used, the space they filled but also the shapes (fan patterns, circles or loops) they drew. Molly drew less and less as she grew older, and we found a significant difference between drawings produced in winter, when orang-utans were kept inside and had less activity, and those produced during other seasons. Our results suggest that the drawing behaviour of these five orang-utans is not random and that differences among individuals might reflect differences of styles, states of mind but also motivation to draw. These novel results are a significant contribution to our understanding of how drawing behaviour evolved in hominids.

**Keywords:** Comparative cognition, scribbles, evolutive anthropology, art, aesthetics

**Introduction**

The origins of drawing behaviour in *Homo sapiens* remain a subject of debate among researchers. The oldest human abstract drawing was recently discovered and is around 73,000 years old (Henshilwood et al. 2018), and the oldest figurative drawing (depicting a wild pig) is around 45,000 years old (Brumm et al. 2021). Given the fact that changes occur gradually in different species, it is interesting to investigate drawing behaviour in our neighbouring species in order to understand its origin. However, since the human species is limited to *sapiens*, the closest species for this study are the great apes, with whom we share many characteristics. But do these related species exhibit drawing behaviour? And if so, how? Drawing behaviour has been studied in non-human primate species including chimpanzees (*Pan troglodytes*), gorillas (*Gorilla gorilla*), orang-utans (*Pongo pygmaeus*), capuchin monkeys (*Cebus apella*) and rhesus macaques (*Macaca mulatta*) (for a review, see Martinet & Pelé, 2020). Nowadays, drawing is increasingly proposed as an enrichment activity for captive great apes in zoological parks and research institutes. The monkeys and apes are free to use the material at their disposal, and are not constrained or conditioned to show this behaviour. This provides a perfect opportunity to collect drawings by non-human primates and allows comparative studies between hominids.

This study is based on 1,433 drawings recovered from five orang-utans (*Pongo pygmaeus*) at Tama Zoological Park in Japan, where caretakers regularly proposed drawing activity to the apes. Orang-utans (*Pongo sp.*) are phylogenetically close to humans, with 96.96% of common genetic heritage (Schwartz 1984). This makes them ideal candidates to help us understand the origins of drawing. They demonstrate highly developed cognitive abilities (Dufour et al. 2009; Damerius et al. 2019a) and tool use (Bardo et al., 2017; Mendes et al., 2007; Lethmate, 1982), that are both at least comparable to those observed in chimpanzees. They are also well known for their curiosity (Damerius et al. 2019b). The wrists of orang-utans are more flexible than those of chimpanzees, making it easier for them to draw as they can bend their hands backwards (Mackinnon 1974). In our drawings database, the daily number of drawings varies significantly between individuals, showing that they had different levels of motivation to draw. This observation leads us to a first question: Do orang-utans show differences in their drawing/marking behaviour? More specifically, it is interesting to note that one female, called Molly, drew almost 1,300 drawings in her last five years of life. She had the opportunity to draw regularly from her 54th birthday onwards (in 2006, at Tama Zoo). Moreover, the colours and page filling in Molly's drawings might be affected by aspects of her daily life such as the identity of her caretaker and events in her environment (Hanazuka et al., 2019). This leads us to a second question: Is there any temporal evolution in her drawing behaviour?

Previous studies on captive orang-utans and other primates showed that they will continue to draw even in the absence of rewards (Boysen et al. 1987; Tanaka et al. 2003; DeLoache et al. 2011; Hanazuka et al. 2019). These findings are consistent with the Gestalt principle found in young children, which links the scribbling activity to a discovery of motor play activity (Casti 2016). Like in humans, spontaneous drawings indicate an intrinsic interest in exploratory and manipulative play for captive non-human primates (Casti 2016; Hanazuka et al. 2019). Moreover, when tracks have already been drawn on the paper, further scribblings are added, suggesting that visible tracks have some kind of reinforcing value (Tanaka et al. 2003). Schiller (1951) worked with an 18-year-old female chimpanzee called Alpha, who showed a keen interest in drawing. In order to study figure formation and position, different stimuli (squares or scribbles) were drawn on paper by researchers. In terms of figure production, the results showed that Alpha mainly used two types of strokes: short dashes, and almost parallel broad zigzag strokes (also referred to as the *fan pattern*). Findings for the placement of figures reveal that when a single figure was positioned off-centre, Alpha drew in the largest open space, producing what Schiller calls a 'sort of balance between her markings and the presented figure'. A similar experiment by Morris (1962) with a one-year-old male chimpanzee named Congo showed similar results: three-quarters of the 40 free drawings on blank sheets of paper showed marks on spaces that had previously been empty, and half of the drawings featured marks that were concentrated at the centre of the paper. Although these results are only descriptive (like those found for Alpha), they do seem to support Schiller's findings concerning the strong tendency to mark a central figure and to position marks in the blank space opposite an offset figure, as well as the inclination to simply enjoy scribbling. Smith (1973) was the first to use quantitative methods in order to analyse chimpanzee drawings. All his results were consistent with the previously cited findings. Later, Boysen et al. (1987) continued the stimulus-drawing test in chimpanzees by presenting 18 different figures to three chimpanzees. Like Smith (1973), Boyson attained a point where the presence of any stimulus figure on the page elicited more centralised markings than in cases where the sheet of paper was blank. The notion of the centre of a page therefore seems to be an important point to take into consideration when studying the emergence of drawing capacities.

Kinematic aspects could be seen as precursors of a graphic representation (Adi-Japha et al. 1998). Yet drawings by non-human primates include different types of marks, such as the straight lines, curves, loops or hook-like strokes observed in drawings by chimpanzees (Tanaka et al. 2003). Although chimpanzees only develop the skill to use a mark-making instrument at the age of 20–23 months, these marks can be observed before the age of 13–23 months through the use of touch-screens. In a comparative study, a set of 396 pictures was collected, made up of 40 drawings by chimpanzees, 153 by gorillas, 146 by orang-utans, and 57 non-figurative drawings by children up to and including the age of four (Zeller 2007). Zeller (2007) noticed that the main features of orang-utans' patterns were diagonals, arcs, and curvilinear designs; those of gorillas contain mainly arcs and open curving lines, and a very high proportion of dots. In contrast, chimpanzees' drawings were mainly characterised by the use of straight and jointed lines. Unlike Tanaka, Zeller noted that orang-utans were the only apes that can use a closed loop or circular pattern (Zeller 2007). This is the most difficult pattern to produce, because it requires high levels of motor control. Another important point concerns the use of colours. Findings in children have shown that their choice of colours reflected their emotional state at that time (Crawford et al. 2012), and it seems that this may be the case in individuals of other species, as with the case of the orang-utan called Molly (Hanazuka et al. 2019). These results support the hypothesis that there is a choice in the use of colour, the type of strokes and the use of space, and that drawings 'do not result from a totally random scribbles' (Zeller 2007; Martinet et al. 2021). Despite their non-figurative nature, drawings produced by apes could therefore provide a great deal of meaningful information.

In this study, we analyse 790 drawings (selected from a total of 1,433 drawings) drawn by five orang-utans (Figure 1) and compare them at an inter-individual level. Indeed, a study of five chimpanzees by Morris (1962) suggests that inter-individual differences exist in drawing and that personality may have an impact on the way marks are distributed. We analyse drawings at an individual level in order to assess temporal changes in this behaviour. We expect to find preferences pertaining to the use of colours, as well as a trend of centering drawings, a use of curved strokes and circles inside the drawings, and possibly a number of differences in marking behaviour between individuals. We also expect possible temporal changes in the drawings of one individual, Molly, who drew almost 1,300 drawings from 2006 to 2011. To our knowledge, this is the first study analysing such a large number of drawings by orang-utans, and indeed by non-human primates in general.

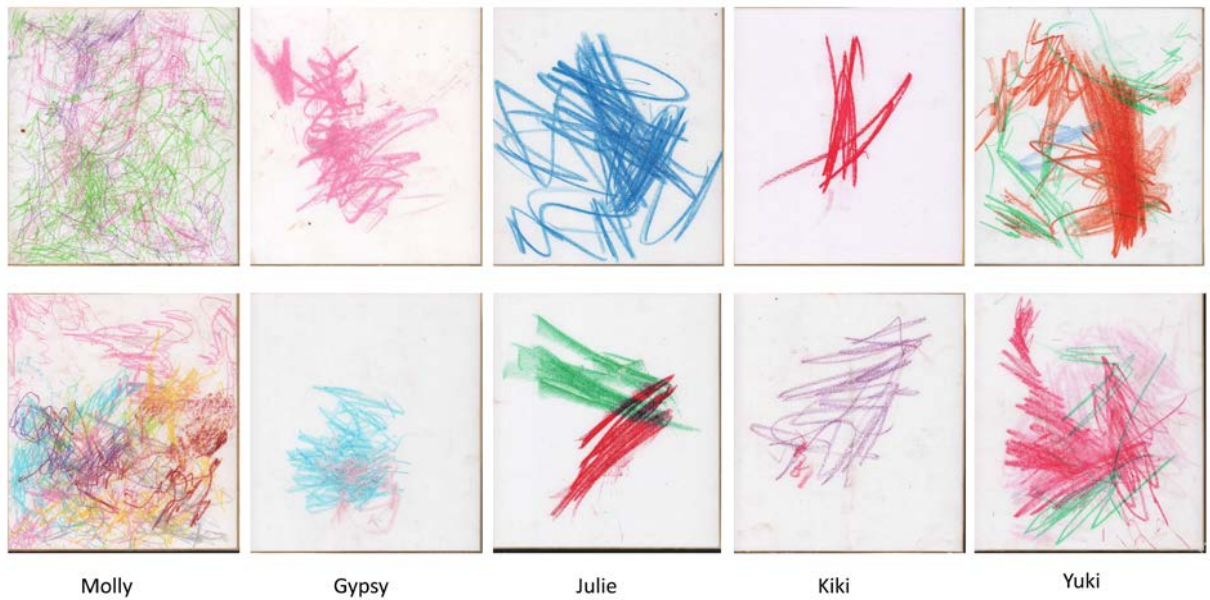

Figure 1: Examples of two drawings for each of the five orang-utans.

**Material & Methods**

*Subjects and collection of drawings*

One thousand four hundred and thirty-three drawings by five female orang-utans at Tama Zoological Park in Tokyo were collected by the caretakers from 2006 to 2016 (Table 1). We analysed 749 drawings (26/26 drawings by Gypsy, 16/16 by Julie, 32/32 by Yuki, 60/60 by Kiki, and 656/1299 by Molly). White, high-quality paperboard (272 x 242 mm) and 16 different coloured crayons were proposed to the orang-utans as permanently available enrichment. Several paperboards were simultaneously provided to orang-utans to avoid competition, but no conflict was observed given the small number of drawings for four of the five individuals. The drawing activity, in the screened restroom, was not part of a research protocol implemented at the zoo but was for enrichment (Sueur and Pelé 2019). The drawings were recovered for analysis after the drawing activities, meaning that we had no control over the methodology. Individuals did not provide the same number of drawings each year, suggesting different motivations between apes and over time. We pseudo-randomly selected 656 drawings by Molly to obtain approximately equal numbers for each three-month period from 2006 to 2011 (about 30 drawings for each of the 19 periods; Molly died in 2011).

Table 1. Information on individuals.

| Name | Date of Birth | Birthplace | Number of drawings | Percentage of drawings | Stay at Tama Zoo | Drawing period |
|---|---|---|---|---|---|---|
| Molly | 01/01/1952 | Wild | 1299 | 91 % | 2005 – 2011 | 2006 – 2011 |
| Gypsy | 01/01/1955 | Wild | 26 | 2% | 1958 – 2017 | 2007 – 2014 |
| Julie | 06/05/1965 | Captivity | 16 | 1% | 2005 – Today | 2009 – 2014 |
| Yuki | 01/01/1970 | Wild | 32 | 2% | 2008 – 2015 | 2008 – 2010 |
| Kiki | 21/10/2000 | Wild | 60 | 4% | 2007 – Today | 2010 – 2016 |

*Ethics note*

The orang-utans were housed in social groups in outdoor and indoor enclosures with environmental enrichments. Food and water were supplied ad libitum. The Tama Zoological Park Ethics Board approved this non-invasive behavioural study, which complied with the Code of Ethics for Japanese Association of Zoos and Aquariums.

*Data collection*

To ensure that our observations were as accurate as possible, we used Gimp 2.10.22 software to apply a 10*10 grid (Casti 2016) to every drawing. Torn sheets were not analysed due to the risk of missing variables. Two observers (GT and AL) made the measurements with a correlation of 0.964≲ $r$ ≲1 for all variables (Figure S1 in Supplementary Material). A third observer (MP) then confirmed the qualitative variable measurements.

*Quantitative variables*

We first looked for non-exclusive quantitative variables (Figure S2): (1) the coverage rate, defined as the number of cells containing one or more strokes out of the total number of cells, (2) the overlap rate, defined as the number of cells containing strokes of at least two different colours that overlap, divided by the coverage rate and multiplied by 100, (3) the solid colour rate, defined as the number of drawing grid cells that were covered at rates of 50% or more, divided by 100, and (4) the distance to the centre, defined as the absolute distance between the centre of an ellipse surrounding the design and the centre of the grid. Three other indices were measured via Gimp 2.10.22 software, namely (5) the number of colours used, (6) the mean deviation of the colour spectrum and (7) the standard deviation of the colour spectrum.

*Qualitative variables*

We also extracted data from every drawing (8) for the predominant colour used by the individual and the drawings shapes, as defined in previous studies (Kellogg, 1969; Tanaka et al., 2003): (9) fan patterns, (10) circles, (11) triangles and (12) loops (Figure S3). These indices are not exclusive, as several can be found on the same drawing. A fan pattern is a stroke making at least three round trips of angles ≤ 45°. A loop is a curved stroke forming a single distinct angle where it intersects itself.

A circle is a curved stroke intersecting in itself without distinct angles. A triangle is a flat loop with distinct angles.

*Statistical analysis*

We conducted two types of analyses: a comparative study between individuals and a longitudinal study of data for Molly. For both, we first checked for high correlations between our drawing variables using the *chart.Correlation()* function of the R package *PerformanceAnalytics* (Peterson and Carl 2020). Our correlation chart revealed no strong correlation between our quantitative variables (Figure S4). A principal component analysis (PCA) was then carried out with the R package *FactoMineR* (Lê et al. 2008) in order to reduce our 12 variables and group them into various dimensions that were then further interpreted from a biological perspective. A PCA with Varimax rotation was also carried out (Kaiser 1958) but did not explain more variance than classical PCA did. For the comparative study between individuals, the coordinates of each drawing in each dimension were used to compare individuals two by two during comparisons of means for each dimension with the functions *kruskal_test()* and *pairwise.wilcox.test()* from the R packages *coin* (Hothorn et al. 2008) and *stats* (R Core Team 2020).

For the longitudinal study of data for Molly, we carried out another PCA. Here, we wanted to study the effect of both 3-month periods (N=19 periods) and seasons (N= 4 seasons) on Molly's drawing behaviour. We then applied a multifactorial linear model (LM) for each dimension of our PCA using the function *lm()* from the R package *car* (Fox and Weisberg 2019). The potential collinearity between our two predictor/predictive variables was tested by the calculation of the variance inflation factor VIF from the R package *car*. These diagnostics revealed a VIF of < 1.1 for both predictors, indicating that there was no notable problem of collinearity. *p*-values for LM were calculated via Monte Carlo sampling with 10,000 permutations, using the function *PermTest()* of the R package *pgirmess* (Giraudoux 2018). Permutation tests for LM were well adjusted for moderate sample size and did not require normal distribution of model residuals (Good 2005).

Pairwise post hoc comparisons for significant LMs were carried out with the function *pairwisePermutationTest()* from the R package *rcompanion* (Mangiafico 2021). α levels (0.05) were Benjamini-Hochberg corrected. Finally, we analysed the main colour used by orang-utans with the function *chisq.test()* of the R package *stats*. We only reported only differences where *p* < 0.05. All statistical analyses were done with R, version 4.0.3 (R Core Team 2020).

**Results**

***Comparative analyses between individuals***

On average, the coverage rate of the drawings by the five orang-utans was 50±30%, the overlap rate was 20±30% and the solid colour rate was 10±20%. As regards the average number of colours used and shapes drawn, the five orang-utans used 3.0 ± 1.8 colours and drew 1.8 ± 2.0 fan patterns, 0.0 ± 0.2 circles, 0.2 ± 0.6 triangles, and 0.7 ± 1.2 loops per drawing. The colour spectrum had a mean of 0.8 ± 0.1 and a standard deviation of 0.1 ± 0.1. Finally, strokes were applied 33.7 ± 25.9 mm from the centre, on average. Variations of each metric per individual are presented in Figure S5.

The three dimensions retained in the PCA had an eigenvalue above 1 and described 63.5% of the explained variance of the dataset. Each metric showed a higher loading in one dimension than in the two others (Table 2, Figure 2). Thus, the first dimension of the PCA (eigenvalue = 3.86, variance = 35.2%) was mainly explained by the filling variables: the recovery rate, the overlap rate, the solid colour rate, the number of fan patterns, the number of colours and the distance to the centre. Colour variables applied to the second dimension (eigenvalue = 1.92, variance = 17.5%), namely the standard deviation and the mean of the colour spectrum. Finally, geometrical shapes were associated with the third dimension (eigenvalue = 1.19, variance = 10.8%), i.e. the number of triangles, loops, and circles.

Table 2. Loadings of the metrics on the three PCA dimensions of our dataset (five orang-utans). Bold values indicate the dimension in which each variable is retained.

|  | Dim.1 | Dim.2 | Dim.3 |
| --- | --- | --- | --- |
| Coverage rate | **0.84** | 0.34 | -0.02 |
| Overlap rate | **0.86** | 0.17 | -0.15 |
| Number of colours | **0.72** | 0.32 | -0.26 |
| Fan patterns | **0.76** | -0.12 | -0.05 |
| Circles | 0.10 | 0.14 | **0.50** |
| Triangles | 0.08 | 0.10 | **0.73** |
| Loops | 0.33 | 0.31 | **0.50** |
| Colour mean | -0.39 | **0.79** | -0.14 |
| Std. deviation of colour mean | 0.38 | **-0.82** | 0.15 |
| Distance to centre | **-0.61** | -0.31 | -0.02 |
| Solid colour rate | **0.74** | -0.37 | -0.07 |

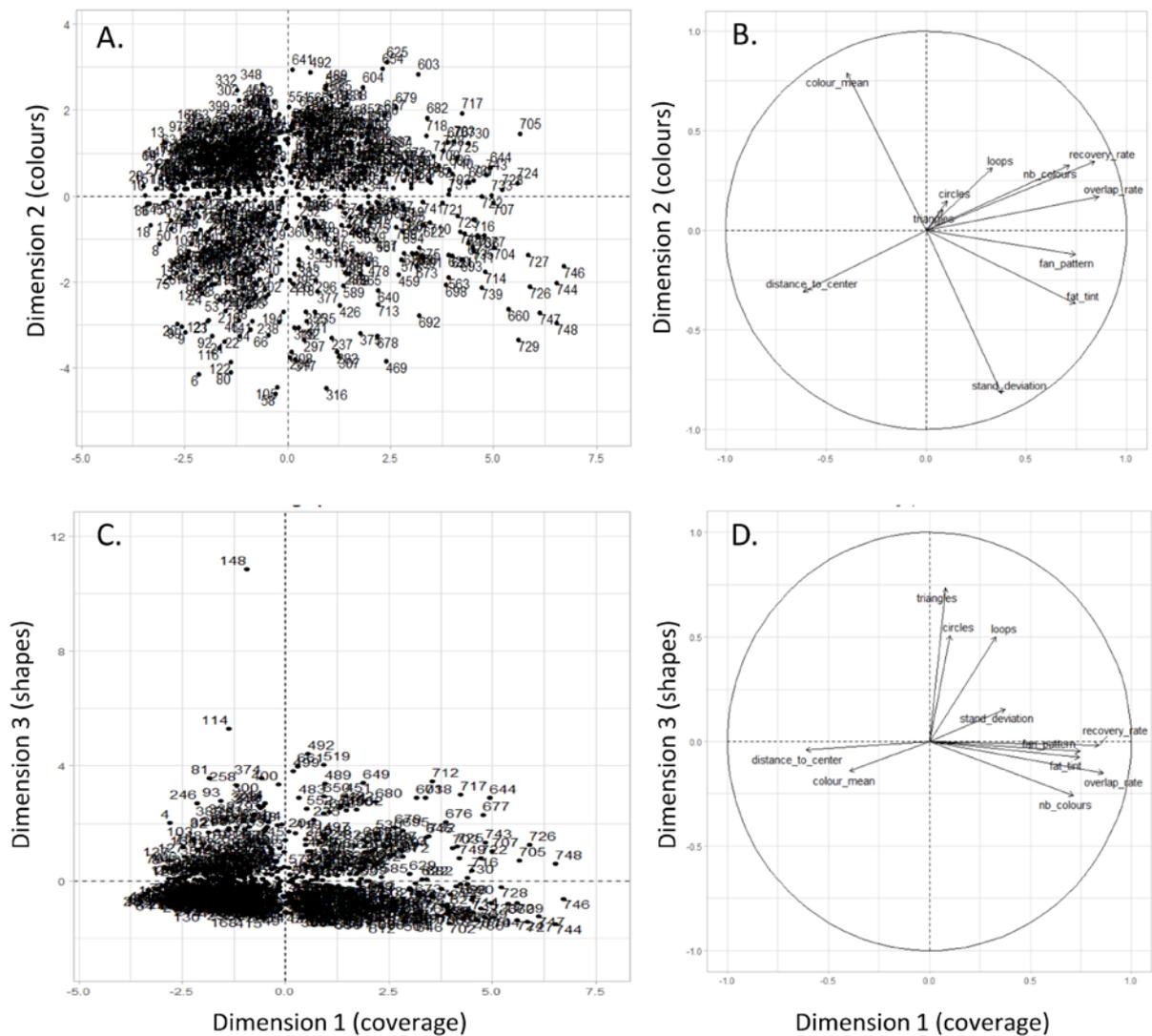

Figure 2: Distribution of data (each point is a drawing) and of variables (each arrow is a variable) respectively for Dimension 1 and Dimension 2 (A., B.) and for Dimension 1 and Dimension 3 (C., D.). The arrow size indicates the correlation with the dimensions.

There were significant differences between the five individuals in the first dimension (Kruskal-Wallis: $\chi^2$ = 50.31, $p < 0.001$). Pairwise comparisons revealed that Molly had higher values than Gypsy and Kiki. Kiki also differed from Julie and Yuki (see pairwise statistics in Table S1, Figure 3a). There were significant differences between orang-utans in the second dimension ($\chi^2$ = 150.48 $p < 0.001$). Pairwise comparisons revealed that Molly had significantly higher values than all other individuals. Julie had lower values than Kiki and Yuki (Table S1, Figure 3b). There were significant differences between individuals in the third dimension ($\chi^2_2$ = 44.51, $p < 0.001$) with differences between Gypsy and Molly

(Table S1, Figure 3c). Statistical analyses for each of the 12 metrics are given in the Supplementary Material section (Table S2-4, Figure S5). The metrics of Dimension 1 are all different between individuals ($\chi^2_2 > 11.53$, $p < 0.02$) with data for Molly being globally different from other individuals (more recovery, more overlap, more colours, closer to the centre) and Kiki showing different results to those of Yuki. Indeed Kiki presented the lowest values for recovering, overlapping and colours whilst showing the second highest values after Molly. Although the metrics of the second dimension also revealed interindividual differences ($\chi^2_2 > 40.08$, $p < 0.001$), the pairwise comparisons yielded more restricted results with only Molly showing very low values and Kiki showing very high values of standard deviation of the colorimetric profile. This indicates that Molly was filling the sheet and showing less contrast, whilst Kiki's drawing showed high contrast due to the few but strong marks on the paper (see Figure 1). Finally, the only difference between individuals in Dimension 3 concerned the number of loops ($\chi^2_2 = 25.07$, $p < 0.001$), with Molly drawing more loops than Kiki.

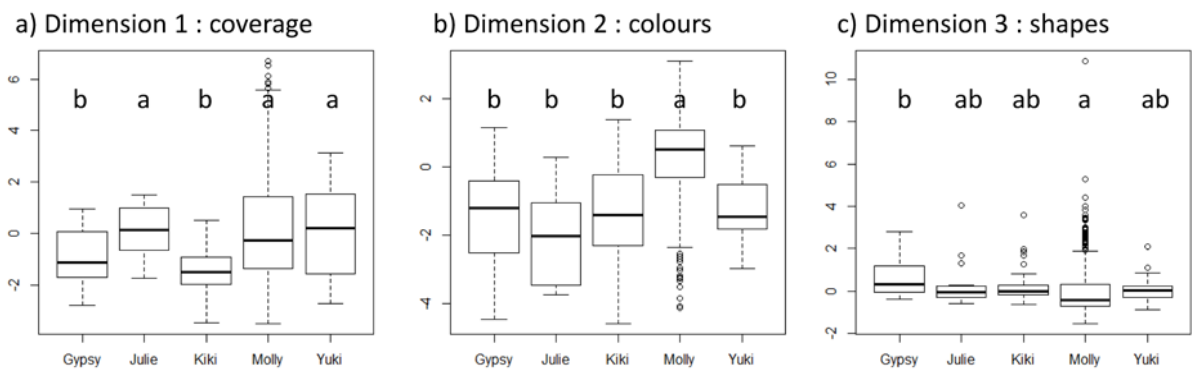

Figure 3: Boxplots showing the differences between individuals in each of the three drawing dimensions. Individuals having different letters show statistical differences.

The main colour found in each drawing revealed a non-random and significant difference in the colours used between individuals (Figure S6). For Kiki ($\chi^2 = 62,647$, $p < 0.001$) and Molly ($\chi^2 = 289.07$, $p < 0.001$), the main colour in the drawings is green (representing 27.45% and 22.6% of the drawings, respectively). For Gypsy ($\chi^2 = 41.72$, $p < 0.001$), Julie ($\chi^2 = 29$, $p = 0.002$) and Yuki ($\chi^2 = 53.2$, $p < 0.001$), the main colour in the drawings is red (representing 40%, 33.3% and 36.67% of the drawings, respectively).

**Longitudinal changes in Molly's drawings**

The number of drawings by Molly over the six years is shown in Figure S7. Three dimensions were retained in the PCA. Their sum represented 64.3% of the explained variance of the dataset (dimension 1: eigenvalue = 4.26, variance = 38.7%; dimension 2: eigenvalue = 1.60, variance = 14.5%; dimension 3: eigenvalue = 1.21, variance = 11%). Metrics are found in the same three dimensions for

Molly as for the four other individuals (Table S5). The effect of 3-month periods and seasons on the three dimensions of the PCA were calculated. There was a significant effect of both seasons (linear model with permutation test: $p < 0.0001$) and periods ($p < 0.0001$) on the first dimension. The filling behaviour of drawings decreased with the periods (linear regression: t-value = -4.7, p<0.001, $R^2$ = 0.03, F=22.73, Figure 4a). Pairwise comparisons revealed that winter had significantly lower values than all other seasons (p<0.005, figure 4b). There was no significant effect of the seasons or of the periods on the second ($p$ = 0.162 and $p$ = 0.869 respectively) or on the third dimension ($p$ = 0561 and $p$ = 0.769 respectively). More precisely, Molly's drawings differed according to the seasons and time in terms of the coverage rate ($p$ = 0.002 and $p$ <0.0001 respectively), the overlap rate ($p$ = 0.002 and $p$ <0.0001), the solid colour rate ($p$ <0.0001 and $p$ <0.0001), the distance to the centre ($p$ =0.04 and $p$ =0.0002) and the fan patterns ($p$ = 0.045 and $p$ <0.0001), with the winter globally always being different from other seasons. However the number of colours used does not differ according to time (p=0.09) or the seasons (p=0.273). Although Dimension 2 did not show differences for the time and the period, the mean colorimetric profile showed differences according to the season (p=0.009) with higher values in winter (p<0.03) than other seasons, because Molly filled the paper sheet less in winter, thus making these drawings brighter. Dimension 3 did not show any differences according to time and period, but the number of loops decreased with time (p=0.02, z-score = -2.29) and was higher in summer (p<0.045) compared to other seasons. Other metrics did not show any difference according to time or seasons.

The analysis of the main colour qualitative variable revealed a non-random and a significant difference in the main colour used by Molly in the four seasons (Figure S8): winter ($X^2$ = 100.16, $p$ < 0.001), spring ($X^2$ = 90,912, $p$ < 0.001), summer ($X^2$ = 87,385, $p$ < 0.001), and autumn ($X^2$ = 103.27, $p$ < 0.001). Molly preferred green in summer and winter and swapped to pink for spring and autumn. Purple is used more in spring compared to other seasons.

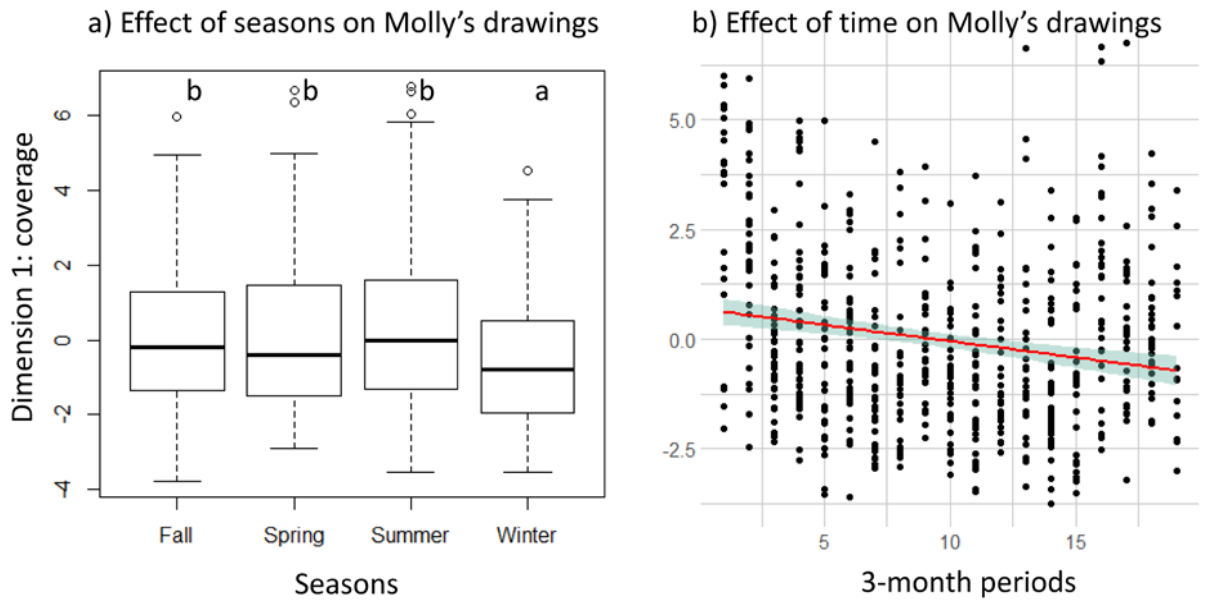

Figure 4: effect of a) seasons and b) time on Dimension 1 (coverage) of Molly's drawings

**Discussion**

We studied drawings by orang-utans in order to consider their cognitive capacities phylogenetically within the framework of other hominoids like humans or chimpanzees. In particular, we looked for differences in the drawing behaviour of five female orang-utans, one of which was in her last years of life. We found differences between individuals and observed temporal changes in Molly's drawings. The principal component analysis reduced the number of variables to three dimensions that had biological, cognitive meaning, namely the filling aspect, the colour or contrast aspect and the shape aspect.

Most of the time, orang-utans used several colours in their drawings. Zeller (2007) found that blue was the most commonly used colour used in the drawings of apes (including orang-utans) and children. The main colour used in our orang-utan's drawings differed between individuals. As orang-utans see colours as we do (Tigges 1963), this difference could reflect either an aesthetic or an emotional preference (Mikellides 2012). In children, colours are linked to emotion (Mikellides 2012), including in their drawing (Crawford et al. 2012). This link with emotion may have an evolutionary origin concerning mate choice and competition (Rowland and Burriss 2017). Molly, who usually used green and pink colours, mainly chose the colour red when another orang-utan was giving birth (Hanazuka et al. 2019). All the individuals drew patterns that were described in Kellogg's children's scribbles classification (Kellogg, 1969). Like children, orang-utans can draw multiple lines (called fan patterns), loops and circles. Orang-utans have more dexterity than other great apes since they can draw curved lines, unlike other great apes (Zeller 2007). Orang-utans have also drawn triangles

(Vancatova 2008). Thus, in a way, orang-utans draw better than other non-human apes. The drawing technique used by orang-utans might also be interesting. Previous studies showed that orang-utans often hold the drawing tool between their fingers. However, one female was regularly observed laying the pastel on the sheet before rolling it with her hand (Vancatova 2008). This behaviour could explain some large fan patterns observed in orang-utan drawings. In many studies, primates demonstrated an ability to draw fan patterns. We can cite the chimpanzees Congo and Bella, and the capuchin monkey Pablo (Morris 1962). Drawings by chimpanzees, human children and orang-utans are therefore comparable. This finding may enrich the phylogeny.

Our comparative study revealed differences in drawing behaviour among orang-utans, especially for the filling dimension. Here, it is not simply a question of one individual standing out from other individuals, but several differences observed between the five individuals: the drawings by Molly are the most complex (more amply filled than pictures by other individuals, with the use of more shapes and colours) followed by those drawn by Yuki. Kiki also shows differences to the other individuals with her simple but strongly marked drawing (i.e. one colour used, pressing hard on the pencil). We can attribute these differences to personality, motivation or even to different cognitive abilities between individuals. Studies in human children suggest that these interindividual differences might be due to varying levels of cognitive skill maturation (Saito et al. 2014) and the different speeds at which children learn to draw (Martinet et al. 2021). According to Willats (2006), there may be an interrelation between a child's stage in their drawing skill development and their increasing comprehension of their living world. Our orang-utans have different life experiences. While Molly, who did more complex drawings, was born in the wild and had lived in two zoological parks and given birth 4 times, Kiki, who produced fewer complex drawings, was born in captivity and quickly moved to Tama Zoological Park where she gave birth to one baby. Molly had also lost her sight in one eye between 1993 and 1996 (personal communications from Mr. Kurotori, Tama zookeeper). These different experiences could perhaps explain the differences found in the way they drew. We can also mention the age difference between Molly, who was 54 years old at the beginning of her drawing period, and Kiki who was the youngest of the study group at just 10 years old. Kiki's minimalist use of colours and space in the paper was evidence of either her drawing style or a lack of motivation or interest in the drawing activity. However, Kiki was not the least experienced of the individuals in terms of drawing, even if she was the youngest: she produced 60 drawings whereas 44-year-old orang-utan Julie produced just 16 drawings.

Data for Molly showed a higher mean and a lower standard deviation of the colour spectrum. This indicates a higher diversity in Molly's drawings, many of which are bright, and have lower levels of contrast. Indeed, although Molly did sometimes fill the sheet, particularly with fan patterns – as described for the chimpanzee Congo (Morris 1962) – many of her drawings were almost empty. Other orang-utans showed smaller patterns, which contrasted with the white background. Moreover, Molly seemed to press less on pencils than other individuals, which explains the lower contrast (lowest standard deviation of the colour spectrum) in comparison to data for other individuals. The drawings of our orang-utans (and especially those drawn by Molly) confirm the findings of Smith (1973), who was the first to report that chimpanzees tended to draw near the centre of the page. Zeller (2007) confirmed this finding for other apes. The inter-individual differences we describe in orangutans drawing in terms of the way individuals draw suggest the existence of different personalities, as first suggested by Morris ( 1962). The wide difference in the number of productions per individual also shows different levels of interest in the activity; this finding confirms previous observations in orang-utans but also in chimpanzees (Boysen et al. 1987; Tanaka et al. 2003; Zeller 2007).

The longitudinal study then demonstrated that Molly's drawings evolved over time and between seasons. She used fewer colours, less space and drew further from the centre as the years progressed. At the end of her life, Molly had fat deposits above her eyes and had to use her hands to lift it out of the way, meaning that she could use her hands less to draw. She also became blind in her left eye, which diminished her visual field. Regarding the difference observed in winter, the results tend to show that Molly's frame of mind changed. The orang-utans in our study group do not go outside in winter, and the weather in Japan is very cold for them. They might become bored, and their motivation may decrease. In this way, our results show that Molly does not systematically use the same main colour in her drawings for all seasons: while she preferred purple as main colour in spring (in 23.4% of her drawings), she used it much less in summer (7.25%), autumn (3.27%) and winter (5.45%). As Hanazuka and his colleagues (2019) hypothesised for non-human primates, drawings could be a window into the internal state of the orang-utan. Moreover, there are no visitors during winter. Evidence that Molly draws more loops and changes the colour she uses in summer could be a cue indicating a good mood due to the weather and the presence of more visitors. Further research is needed to assess whether this number of loops is just the reflect of Molly's motivation or have a more cognitive aspect as symbolism (Gardner and Wolf 1987).

However, this study is limited by the enrichment origin of these drawings, which restrict the possibilities of controlled conditions. Several drawings were covered with stains and were dirty, with some shapes overlapping. It is difficult to be precise and objective when measuring variables such as the shapes present in the drawing or the predominant colour, especially when water had blurred the patterns. In the future, we could therefore use touch-screen tablets instead of paper (Martinet et al. 2021). The use of a tablet requires a genuine trust between the researcher and the animal, established from an early age so as not to put the researcher in danger (Matsuzawa 2017). Drawing on tablets has several advantages such as the extraction of more important information such as temporal data. This method will reveal differences between individuals in terms of order of colours used but also the fractality of the patterns (Martinet et al. 2021). It would also be interesting to study drawings by males to see if these preferences change according to sex, although this may be difficult as males have previously demonstrated less interest in this task (Vancatova 2008). Another question pertains to the part of the body orang-utans use to draw. Indeed, some apes use their hands while others use their mouth (Vancatova 2008).

Overall, the fact that orang-utans draw freely without constraint shows that they are capable of doing so. It would be interesting to study these individuals further and also to study mother-offspring pairs in order to identify a possible transmission of the motivation or ability to draw. Cultural transmission of habits has been shown for example in chimpanzees (Whiten et al. 1999), and in orang-utans (Schaik et al. 2003). This type of study would contribute to our understanding of the origins of drawings in humans.

**Acknowledgements**

We are grateful to the Tama Zoological Park in Japan for the providence of orang-utans drawings. This project has received financial support from the CNRS through the MITI interdisciplinary programs and from a KAKENHI program.

**Supplementary material for Interindividual and intraindividual differences in orangutans drawings**

Marie Pelé[1], Gwendoline Thomas[2], Alaïs Liénard[2], Shimada Masaki[3], Cédric Sueur[4,5]
1: Anthropo-Lab, ETHICS EA7446, Lille Catholic University, Lille, France
2 : Université Sorbonne Paris Nord - UFR LLSHS, Paris, France
3 : Department of Animal Sciences, Teikyo, University of Science, Uenohara, Yamanashi, Japan
4: Université de Strasbourg, CNRS, IPHC UMR 7178, Strasbourg, France
4 : Institut Universitaire de France, Paris, France

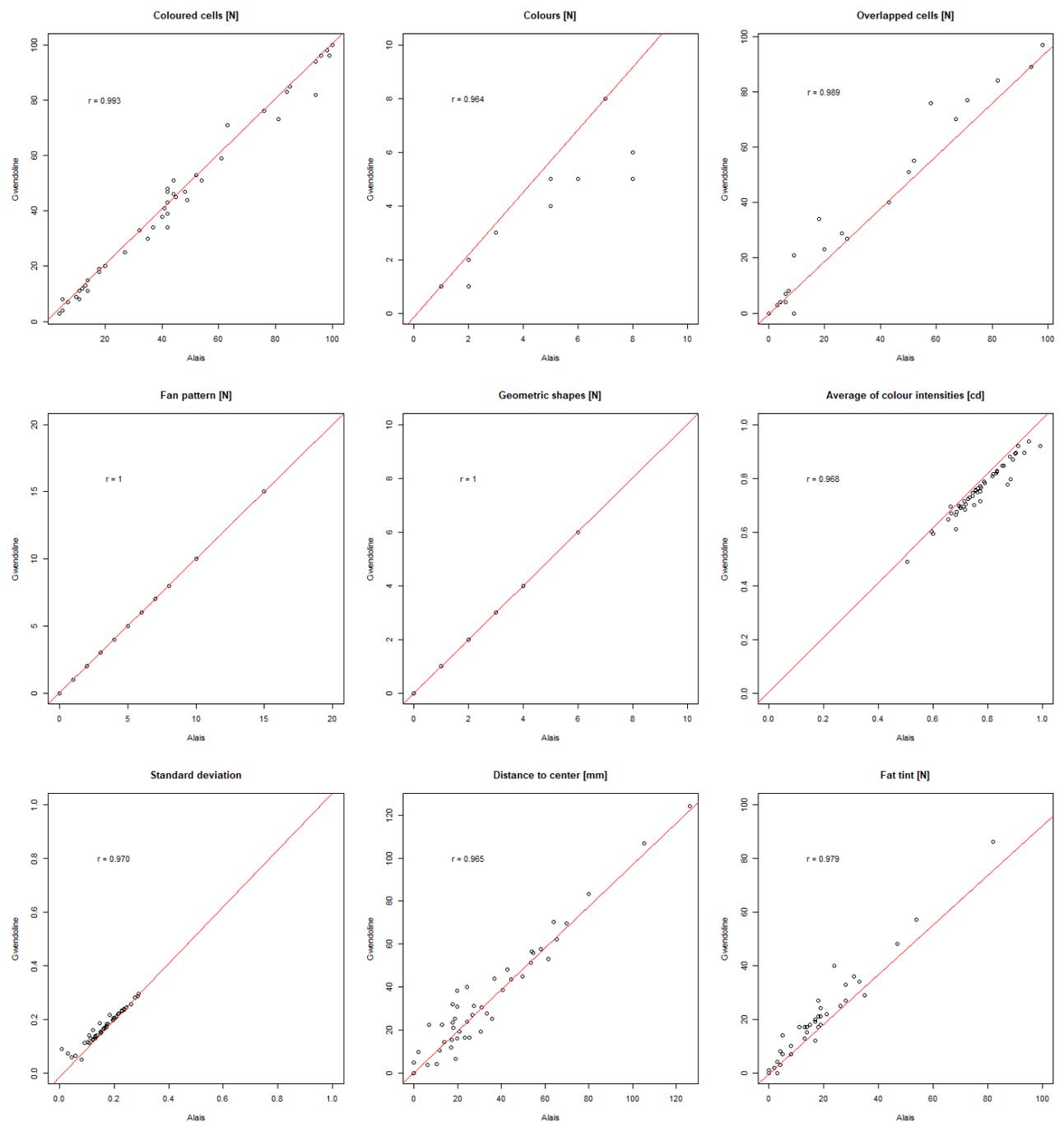

Figure S1. Inter-observer correlations for quantitative variables in the classical analysis.

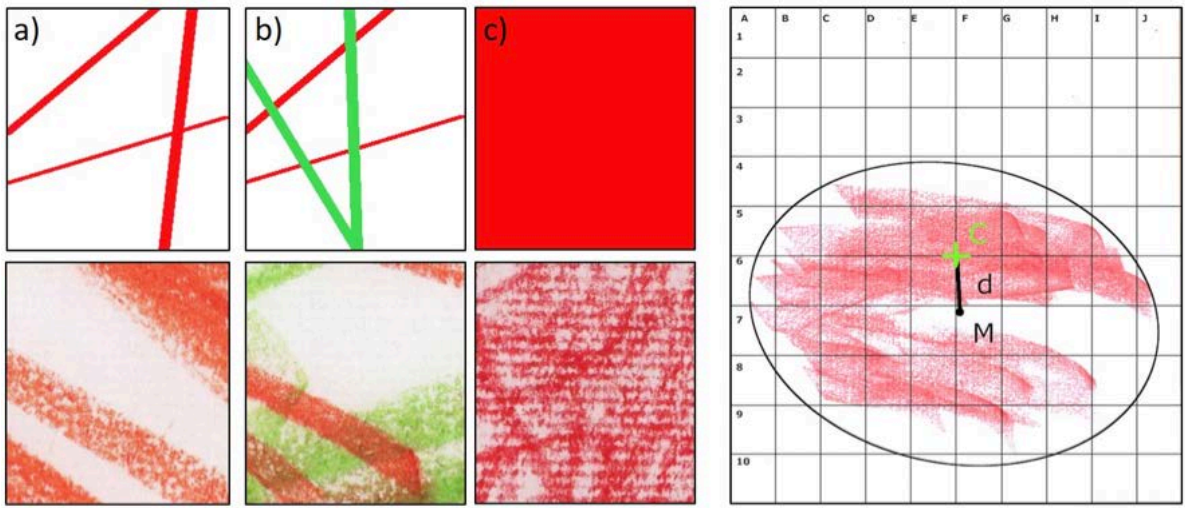

Figure S2: On the left, schematic (top) and actual (bottom) examples of a) a covered cell, b) an overlapped cell and c) solid colour rate. On the right, calculation of the distance to the centre. C is the centre of the paper sheet, M is the centre of the drawing ellipse, and d is the diameter.

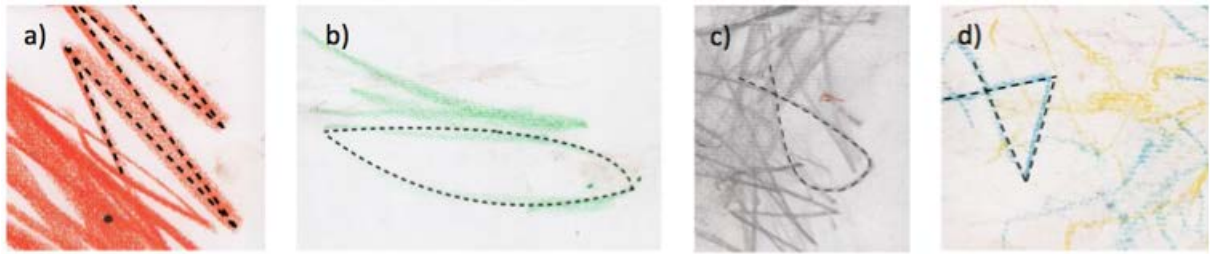

Figure S3: examples of a) a fan pattern, b) a circle, c) a loop and d) a triangle.

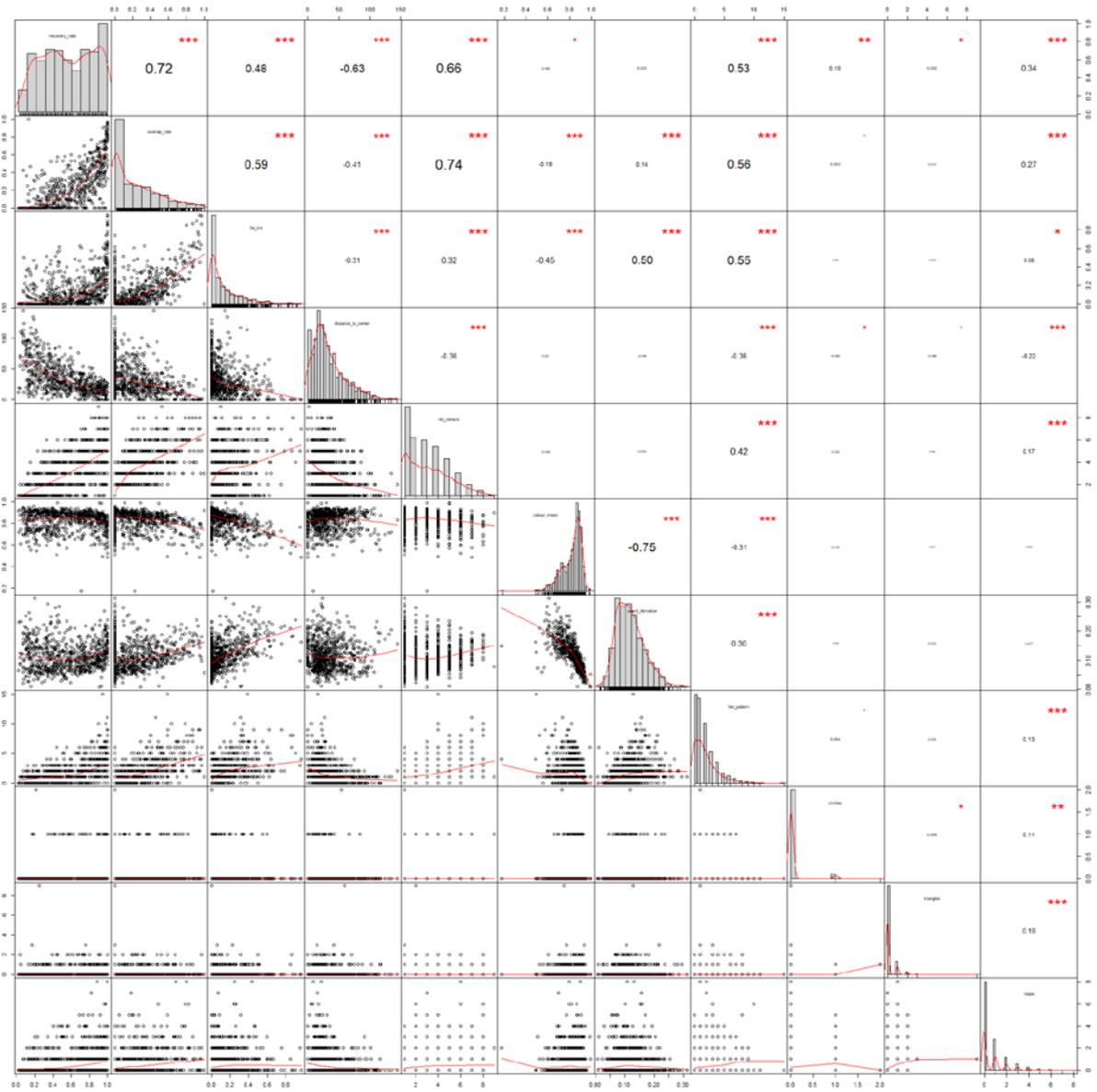

Figure S4. Correlation chart of our quantitative variables. *:p<0.05, **p<0.01,***:p<0.001. The number indicates the correlation.

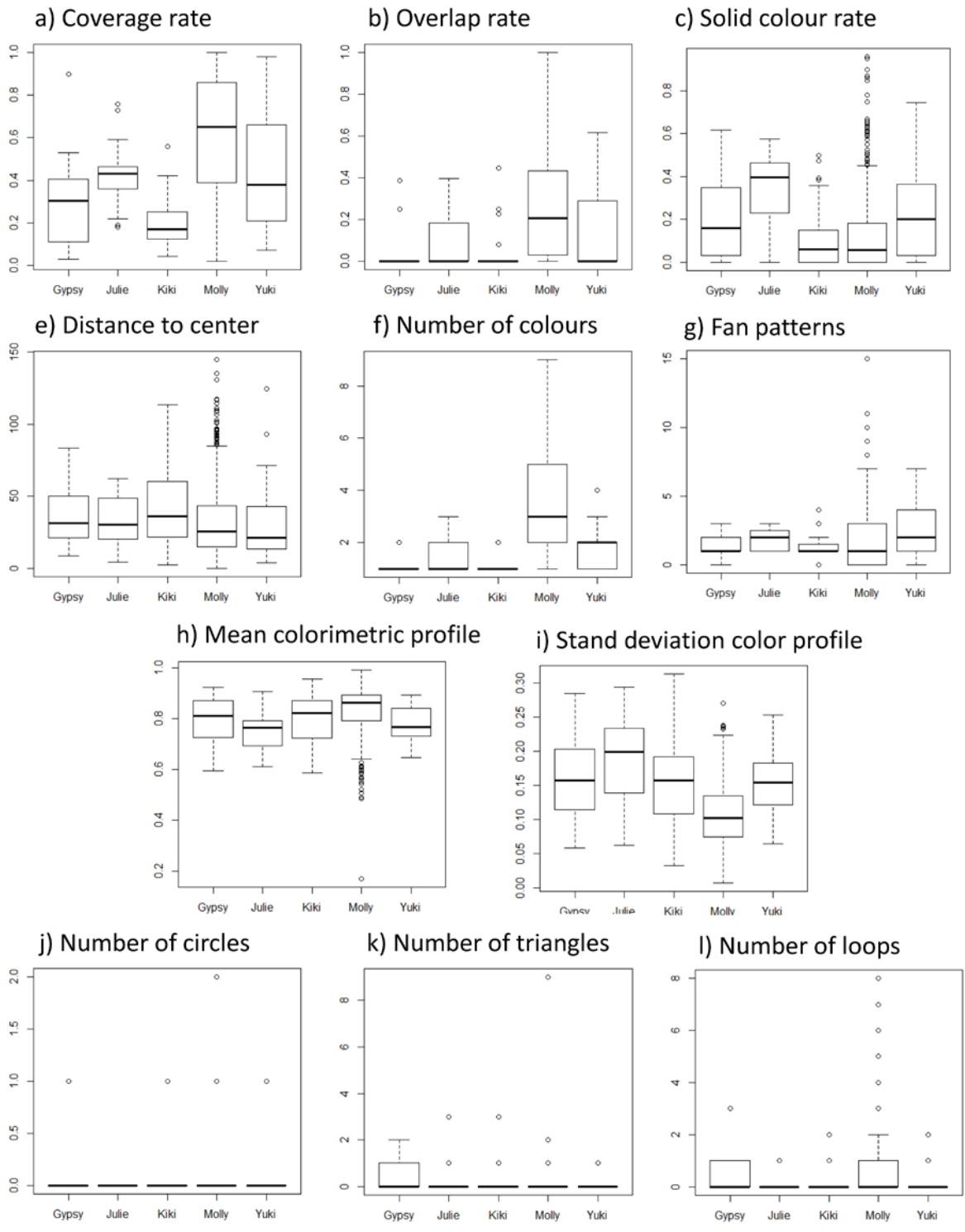

Figure S5: Boxplots of each drawing metric per individual

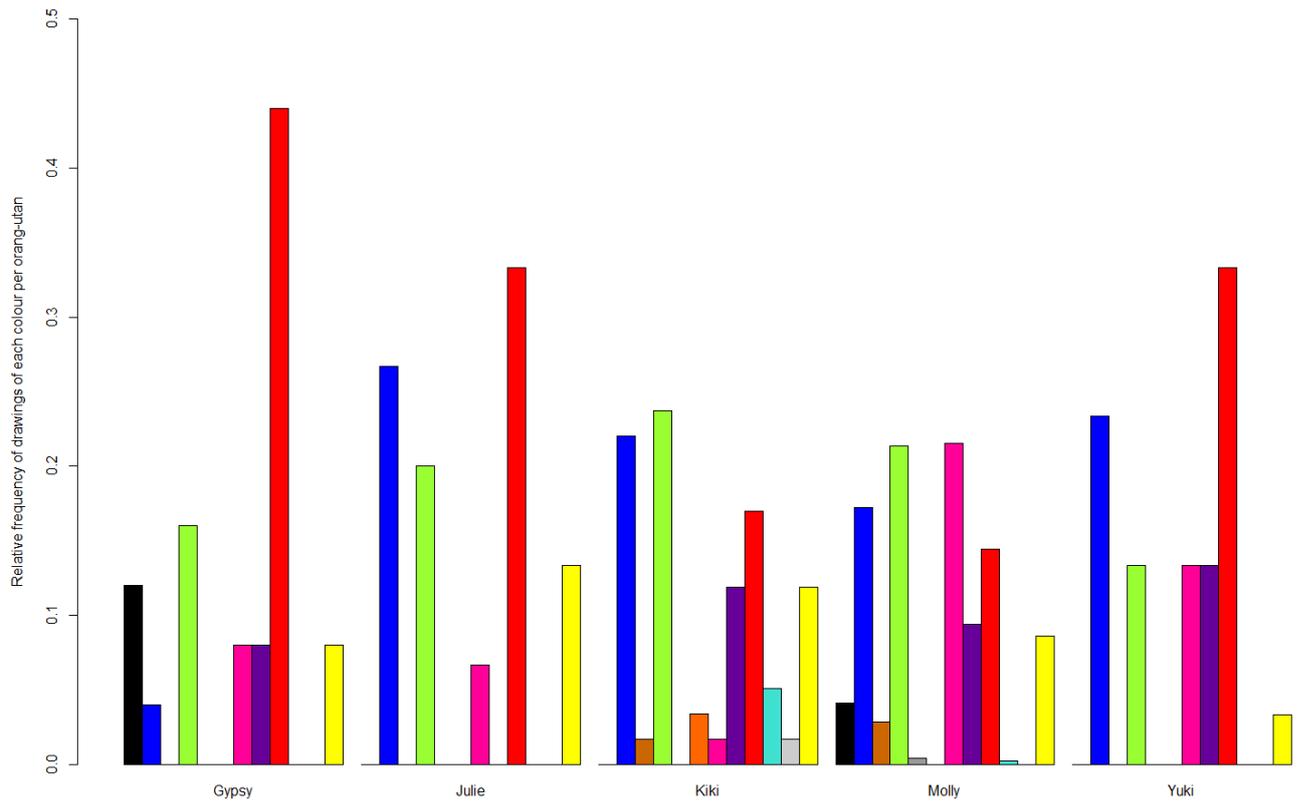

Figure S6. Frequency of drawings per main colour per individual.

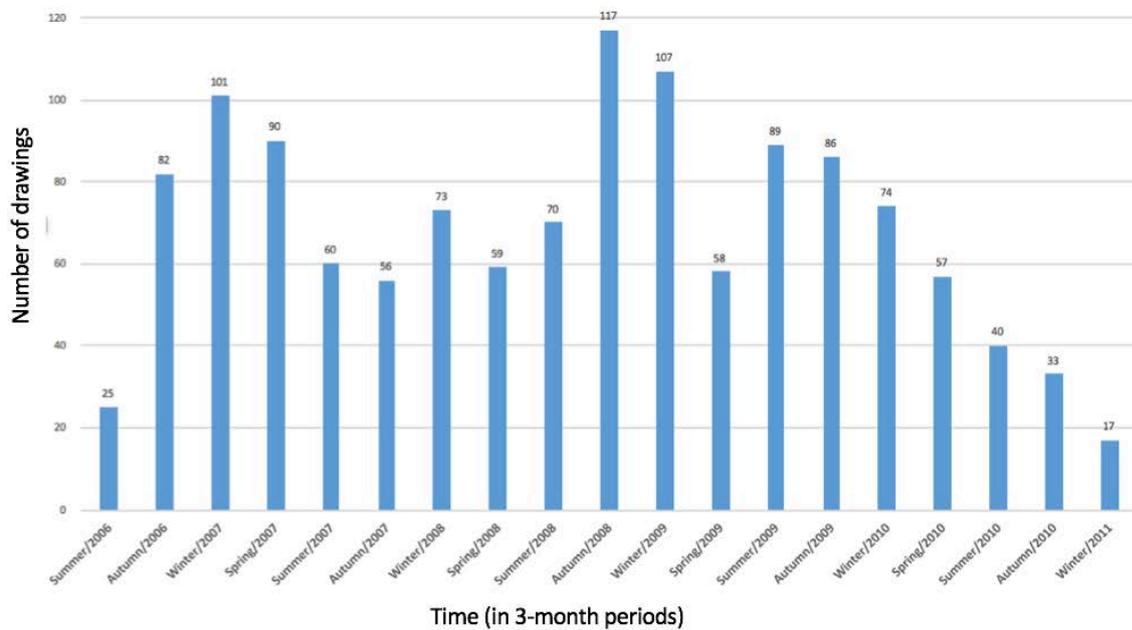

Figure S7: Number of drawings according to the seasons over the years. The exact number is specified above the bars.

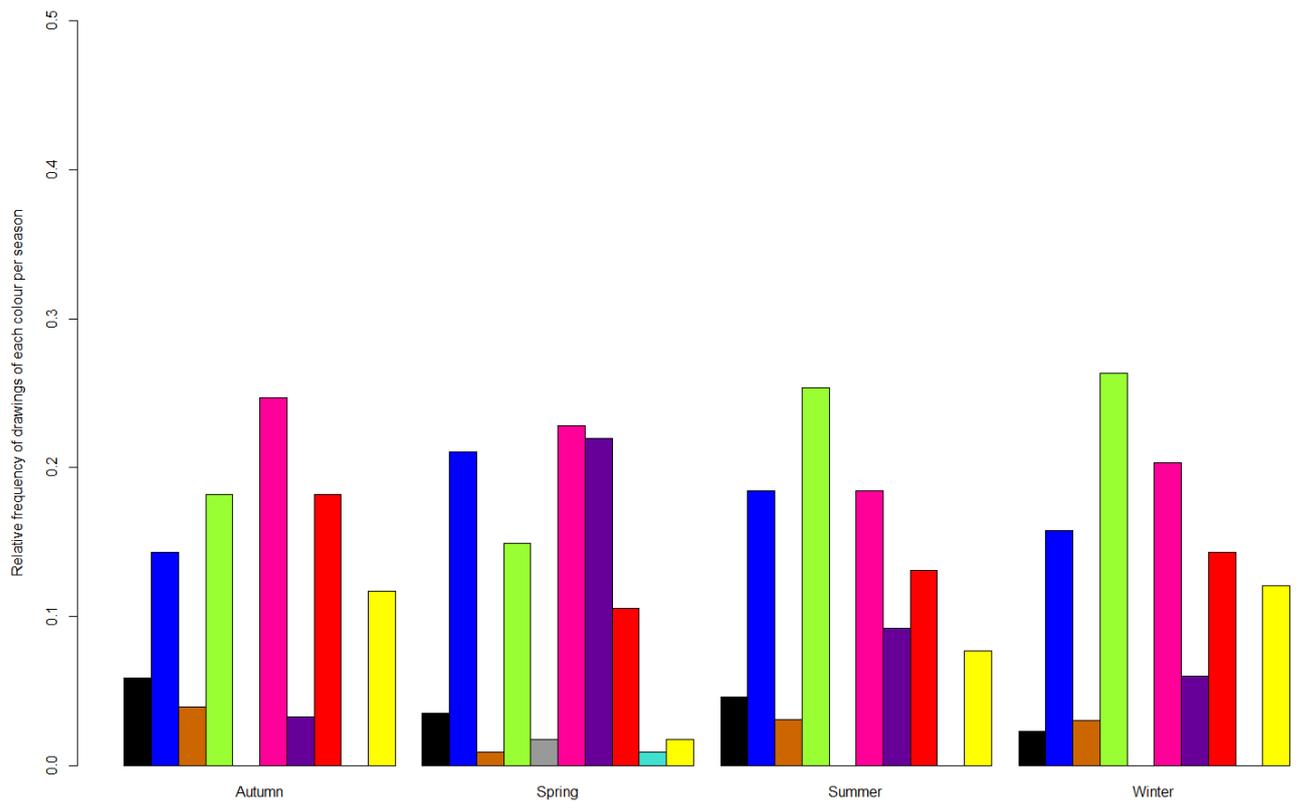

Figure S8. Frequency of Molly's drawings per main colour per season.

Table S1: pairwise comparisons tests (p-values) for each dimension between the five individuals

| **Dim1** | **Gypsy** | **Julie** | **Kiki** | **Molly** |
|---|---|---|---|---|
| **Julie** | 0.1844 | - | - | - |
| **Kiki** | 0.3149 | 0.0152 | - | - |
| **Molly** | 0.016 | 0.9006 | 3.60E-09 | - |
| **Yuki** | 0.1525 | 0.9006 | 0.0025 | 0.8234 |
| **Dim2** | Gypsy | Julie | Kiki | Molly |
| **Julie** | 0.081 | - | - | - |
| **Kiki** | 0.932 | 0.045 | - | - |
| **Molly** | 4.80E-11 | 8.30E-14 | < 2e-16 | - |
| **Yuki** | 0.907 | 0.045 | 0.907 | 5.00E-12 |
| **Dim3** | Gypsy | Julie | Kiki | Molly |
| **Julie** | 0.643 | - | - | - |
| **Kiki** | 0.301 | 0.643 | - | - |
| **Molly** | 0.041 | 0.301 | 0.301 | - |
| **Yuki** | 0.301 | 0.601 | 0.643 | 0.643 |

Table S2: pairwise comparisons tests (p-values) for each metric of Dimension 1

| Coverage rate | Gypsy | Julie | Kiki | Molly |
|---|---|---|---|---|
| Julie | 0.13059 | - | - | - |
| Kiki | 0.17047 | **0.00479** | - | - |
| Molly | **1.50E-08** | **0.01076** | **< 2e-16** | - |
| Yuki | 0.07067 | 0.99024 | **0.00029** | **0.0004** |
| Overlap rate | Gypsy | Julie | Kiki | Molly |
| Julie | 0.482 | - | - | - |
| Kiki | 0.869 | 0.372 | - | - |
| Molly | **8.40E-06** | **0.018** | **5.60E-13** | - |
| Yuki | 0.112 | 0.549 | **0.032** | **0.018** |
| Number of colours | Gypsy | Julie | Kiki | Molly |
| Julie | 0.541 | - | - | - |
| Kiki | 0.931 | 0.541 | - | - |
| Molly | **1.30E-10** | **1.90E-05** | **< 2e-16** | - |
| Yuki | 0.146 | 0.541 | 0.093 | **2.10E-06** |
| Fan patterns | Gypsy | Julie | Kiki | Molly |
| Julie | 0.5247 | - | - | - |
| Kiki | 0.8489 | 0.346 | - | - |
| Molly | 0.329 | 0.9918 | **0.0406** | - |
| Yuki | **0.0406** | 0.3048 | **0.0044** | **0.0406** |
| Distance to centre | Gypsy | Julie | Kiki | Molly |
| Julie | 0.911 | - | - | - |
| Kiki | 0.714 | 0.668 | - | - |
| Molly | 0.714 | 0.912 | **0.035** | - |
| Yuki | 0.714 | 0.912 | 0.271 | 0.912 |
| Solid colour rate | Gypsy | Julie | Kiki | Molly |
| Julie | 0.0212 | - | - | - |
| Kiki | 0.0189 | **3.50E-06** | - | - |
| Molly | **0.0287** | **3.50E-06** | 0.319 | - |
| Yuki | 0.5568 | **0.0461** | **0.0015** | **0.0015** |

Table S3: pairwise comparisons tests (p-values) for each metric of Dimension 2

| Colour mean | Gypsy | Julie | Kiki | Molly |
|---|---|---|---|---|
| Julie | 0.2651 | - | - | - |
| Kiki | 0.8763 | 0.2651 | - | - |
| Molly | 0.1266 | 0.0057 | 0.0057 | - |
| Yuki | 0.5863 | 0.5263 | 0.5863 | 0.0059 |
| Standard deviation of mean colour | Gypsy | Julie | Kiki | Molly |
| Julie | 0.0335 | - | - | - |
| Kiki | 0.5064 | 0.0022 | - | - |
| Molly | 1.40E-07 | 1.20E-11 | 5.60E-11 | - |
| Yuki | 0.7355 | 0.0135 | 0.6962 | 9.40E-08 |

Table S4: pairwise comparisons tests (p-values) for each metric of Dimension 3

| Circles | Gypsy | Julie | Kiki | Molly |
|---|---|---|---|---|
| Julie | 0.88 | - | - | - |
| Kiki | 0.88 | 0.88 | - | - |
| Molly | 0.88 | 0.88 | 0.88 | - |
| Yuki | 0.89 | 0.88 | 0.88 | 0.88 |
| **Triangles** | Gypsy | Julie | Kiki | Molly |
| Julie | 1 | - | - | - |
| Kiki | 0.45 | 0.45 | - | - |
| Molly | 0.45 | 0.54 | 0.45 | - |
| Yuki | 0.45 | 0.45 | 0.89 | 0.65 |
| **Loops** | Gypsy | Julie | Kiki | Molly |
| Julie | 0.5006 | - | - | - |
| Kiki | 0.5006 | 0.835 | - | - |
| Molly | 0.5006 | 0.1012 | **0.0022** | - |
| Yuki | 0.5542 | 0.835 | 0.835 | 0.079 |

Table S5: Loadings of the metrics on the three PCA dimensions of our dataset for Molly. Bold values indicate the dimension in which each variable is retained.

| | Dim.1 | Dim.2 | Dim.3 |
|---|---|---|---|
| **coverage rate** | **0.83259466** | 0.35898791 | -0.03754139 |
| **Overlap rate** | **0.86645016** | 0.11386246 | -0.15141994 |
| **Solid colour rate** | **0.82093692** | -0.20890006 | -0.06820954 |
| **distance to centre** | **-0.59224515** | -0.4847952 | 0.0366672 |
| **Number of colours** | **0.70657478** | 0.21081542 | -0.29585089 |
| **Colour mean** | -0.5250608 | **0.71138628** | -0.22231713 |
| **Standard deviation of mean colour** | 0.59217275 | **-0.65153752** | 0.21292017 |
| **fan pattern** | **0.75009581** | -0.06012067 | -0.05905462 |
| circles | 0.09579288 | 0.18189447 | **0.50532385** |
| triangles | 0.07651598 | 0.16002544 | **0.70546716** |
| loops | 0.30038742 | 0.37499276 | **0.49370317** |